# Exact solutions to the geodesic equations of linear dilaton black holes

**Alan Hameed Hussein HAMO[1], Izzet SAKALLI[1,*]**

[1]Department of Physics, Eastern Mediterranean University, Gazimagosa-KKTC, via Mersin 10 Turkey.

**Abstract:** In this paper, we analyze the geodesics of the 4-dimensional ($4D$) linear dilaton black hole (LDBH) spacetime, which is an exact solution to the Einstein-Maxwell-Dilaton (EMD) theory. LDBHs have non-asymptotically flat (NAF) geometry, and their Hawking radiation is an isothermal process. The geodesics motions of the test particles are studied via the standard Lagrangian method. After obtaining the Euler-Lagrange (EL) equations, we show that exact analytical solutions to the radial and angular geodesic equations can be obtained. In particular, it is shown that one of the possible solutions for the radial trajectories can be given in terms of the WeierstrassP-function ($\wp$-function), which is an elliptic-type special function.

**Key words**: Linear dilaton black hole, Geodesics, WeierstrassP-function.

## 1. Introduction

The motion of test particles (both massive and massless) provides the only experimentally feasible way to study the gravitational fields of objects such as black holes (BHs). Predictions about their observable effects (light deflection, the perihelion shift, gravitational time-delay etc.) can be made, and also compared with the observations [1]. For this reason, the subject of geodesics in the BH spacetimes have always been attracted much attention. Today, there are numerous studies about the geodesics of various BHs in the literature (for instance, one may see [2] and references therein). Recently, studies on the general solution to the geodesic equation in $4D$ spacetimes have considerably increased [3-6].

---

* Correspondence: izzet.sakalli@emu.edu.tr

In this paper, our main motivation is to study the geodesic structure of the LDBH [7,8] whose asymptotic behavior is NAF. This BH arises as an exact solution to the EMD theory [7-9]. One of the most intriguing features of those BHs is that their Hawking radiation (HR) is governed by isothermal processes, which occur at a constant temperature. Namely, while a LDBH radiates, the energy transferring out of the BH happens at such a slow rate that the thermal equilibrium is always maintained. The studies on the LDBHs, which are subject to quantum gravity theory (including thermodynamics), can be seen in [10-18].

In the present study, we study the geodesic motion of a generic test particle (timelike, spacelike and null geodesics) on the LDBH background. To that end, we follow the standard Lagrangian procedure with the Mino proper time [19]. We give analytical expressions for the radial and angular geodesic equations. Particularly, the radial ones are found in terms of both hyperbolic functions and the $\wp$-function [20].

The paper is organized as follows. In Sec. 2, we review the LDBH spacetime and present some of its physical properties. Sec. 3 is devoted to the exact analytical solutions of the geodesic equations in the LDBH background. We draw our conclusions in Sec. 4.

## 2. LDBH Spacetime

In general, the metric of a static and spherically symmetric BH in $4D$ is given by:

$$ds^2 = -fdt^2 + f^{-1}dr^2 + R^2 d\Omega^2, \qquad (1)$$

where

$$d\Omega^2 = d\theta^2 + \sin\theta^2 d\phi^2. \qquad (2)$$

Equation (2) is the line-element of the unit 2-sphere. When the metric functions of the line-element (1) are written in the following forms:

$$f = \frac{1}{r_0}(r-b), \qquad (3)$$

$$R^2 = r_0 r, \qquad (4)$$

the spacetime (1) is called as the LDBH [7,8]. Here, the physical constant parameter $r_0$ is related with the conserved charge of the LDBH: $r_0 = \sqrt{2}Q$ in which the charge $Q$ is a non-zero positive definite physical parameter [7,8]. However, these BHs have no zero charge limit due to the associated field equations coming from the EMD theory. More details about the features of the LDBH can be found in the papers written by Clément et al. [7,8].

It is obvious from Eq. (3) that a LDBH possesses a NAF geometry, and its event horizon is $r_h = b$. For $b \neq 0$, the horizon transparently shields the null singularity at $r = 0$. On the other hand, if one applies the quasi-local mass ($M$) definition of Brown and York [21] to the metric (1) with metric functions (3) and (4), we obtain

$$b = 4M. \qquad (5)$$

In general, the definition of the Hawking temperature $T_H$ [22,23])is expressed in terms of the surface gravity $\kappa$ [24]:

$$\kappa = \sqrt{\left[-\frac{1}{4}\lim_{r\to r_h}\left(g^{tt}g^{ij}g_{tt,i}g_{tt,j}\right)\right]}, \qquad (6)$$

as

$$T_H = \frac{\kappa}{2\pi}. \qquad (7)$$

Using Eq. (6), one can compute the surface gravity $\kappa = \frac{1}{2r_0}$. Thus, $T_H$-value of the LDBH becomes:

$$T_H = \frac{1}{4\pi r_0} \qquad (8)$$

It is obvious from the above expression that the obtained temperature is constant; thus $\Delta T = 0$. So, the HR of the LDBH is made by the series of the isothermal processes.

One can also compute that the geometric scalars (Ricci scalar ($\mathfrak{R}$), full contradiction of the Ricci tensor ($\mathfrak{R}_{\alpha\beta}\mathfrak{R}^{\alpha\beta}$) and Kretschmann scalar ($\aleph$)) of the spacetime [25]:

$$\mathfrak{R} = \frac{1}{2r_0}\left(\frac{1}{r} - \frac{b}{r^2}\right), \qquad (9)$$

$$\mathfrak{R}_{\alpha\beta}\mathfrak{R}^{\alpha\beta} = \frac{1}{4r_0^2}\left(\frac{b^2}{r^4} + \frac{3}{r^2}\right), \qquad (10)$$

$$\aleph = \mathfrak{R}_{\alpha\beta\mu\nu}\mathfrak{R}^{\alpha\beta\mu\nu} = \frac{1}{4r_0^2}\left(\frac{3b^2}{r^4} + \frac{6b}{r^3} + \frac{11}{r^2}\right), \qquad (11)$$

which represents that the curvature singularity is located at $r = 0$.

## 3. Analytic solution to the geodesic equations of the LDBH geometry

In this section, for studying the geodesics of the test particles in the LDBH background, we shall employ the standard Lagrangian method. The corresponding Lagrangian $(L)$ of a test particle in the LDBH geometry is given by

$$2L = -f\dot{t}^2 + \frac{\dot{r}^2}{f} + r_o r\left(\dot{\theta}^2 + sin^2\theta \dot{\phi}^2\right), \qquad (12)$$

where the dot over a quantity denotes the derivative with respect to the affine parameter $\sigma$. The metric condition is in general defined by

$$L = \frac{\varepsilon}{2}. \qquad (13)$$

Here $\varepsilon = (0,-1,1)$ stands for the (null, timelike, spacelike) geodesics. Now, we reconsider the Lagrangian (12) by using the Mino proper time $(\gamma)$ [19] which is governed by the following differential expression for the LDBH

$$d\sigma = \sqrt{rr_0}d\gamma. \qquad (14)$$

Thus, the modified Lagrangian becomes

$$L = -\frac{1}{2}\frac{f}{rr_o}\left(\frac{dt}{d\gamma}\right)^2 + \frac{1}{2rr_o f}\left(\frac{dr}{d\gamma}\right)^2 + \frac{1}{2}\left(\frac{d\theta}{d\gamma}\right)^2 + \frac{1}{2}sin^2\theta\left(\frac{d\phi}{d\gamma}\right)^2. \qquad (15)$$

After applying the EL method, we obtain

$$\frac{d}{d\gamma}\left(-\frac{f}{rr_0}\frac{dt}{d\gamma}\right) = 0 \Rightarrow \frac{dt}{d\gamma} = \frac{r^2 r_0}{\Lambda}\alpha, \qquad (16)$$

$$\frac{d}{d\gamma}\left(sin^2\theta\frac{d\phi}{d\gamma}\right) = 0 \Rightarrow \frac{d\phi}{d\gamma} = \frac{\beta}{sin^2\theta}, \qquad (17)$$

in which $\Lambda = rf$. Besides, $\alpha$ and $\beta$ are the real integration constants. In addition to Eqs. (16) and (17), we have

$$\frac{d^2\theta}{d\gamma^2} = \sin\theta\cos\theta\left(\frac{d\phi}{d\gamma}\right)^2, \tag{18}$$

which is equivalent to

$$\left(\frac{d\theta}{d\gamma}\right)^2 = K - \left(\frac{\beta}{\sin\theta}\right)^2 \tag{19}$$

where $K$ is another integration constant. Finally, with the aid of the metric condition (13), one can derive the radial equation as follows

$$-\frac{1}{2}\frac{f}{rr_0}\left(\frac{r^4 r_0^2}{\Lambda^2}\alpha^2\right) + \frac{1}{2rr_0 f}\dot{r}^2 + \frac{1}{2}\left(K - \frac{\beta^2}{\sin^2\theta}\right) + \frac{1}{2}\sin^2\theta\frac{\beta^2}{\sin^4\theta} = \frac{\varepsilon}{2},$$

$$\Rightarrow -\frac{r^2 r_0}{\Lambda}\alpha^2 + \frac{1}{\Lambda r_0}\dot{r}^2 = \varepsilon - K, \tag{20}$$

$$\Rightarrow \dot{r}^2 = (\varepsilon - K)\Lambda r_0 + (\alpha rr_0)^2.$$

Following the method whose details are given in [26,27], we can make a transformation for the $r$-coordinate as

$$r = \frac{s}{x} + z, \tag{21}$$

where $s = \pm 1$ (i.e., $s^2 = 1$) and $z$ satisfies the following condition

$$\left. (\varepsilon - K)\Lambda r_0 + (\alpha rr_0)^2 \right|_{r=z} = 0. \tag{22}$$

Recalling that $\Lambda = rf$, one can easily verify that Eq. (22) admits

$$z = \frac{\Xi}{\Upsilon}, \tag{23}$$

where

$$\Xi = b(\varepsilon - K), \quad (24)$$

$$\Upsilon = \alpha^2 r_0^2 + \varepsilon - K. \quad (25)$$

Then, Eq. (20) becomes

$$\left(\frac{dx}{d\gamma}\right)^2 = b_3 x^3 + b_2 x^2. \quad (26)$$

where

$$b_2 = \frac{\alpha^2 r_0^2 b}{b - z} = \Upsilon, \quad (27)$$

$$b_3 = \frac{z}{s} b_2 \equiv szb_2 = s\Xi. \quad (28)$$

Equation (26) has two solutions. One of them is

$$x_1 = -\frac{s}{z}, \quad (29)$$

which yields $r = 0$, so that it is a trivial solution. Another solution can be found in terms of the hyperbolic functions as follows

$$x_2 = -\frac{s}{z \cosh^2\left[\frac{\sqrt{b_2}}{2}(\gamma - c)\right]}, \quad (30)$$

where $c$ is an integration constant. Meanwhile it is worth noting that Eq. (26) can be converted to a $\wp$-equation [20] by the following transformation

$$x = \frac{12U - b_2}{3b_3} = \frac{s(12U - b_2)}{3zb_2}, \tag{31}$$

where $U = U(\gamma)$. Then, Eq. (26) transforms to the $\wp$-equation [28,29]

$$\left(\frac{dU}{d\gamma}\right)^2 = 4U^3 - \frac{b_2^2}{12}U + \frac{b_2^3}{216}. \tag{32}$$

which admits two solutions:

$$U_1 = -\frac{b_2}{6}, \tag{33}$$

$$U_2 = \frac{1}{6}\wp\left(\frac{\gamma}{\sqrt{6}} + \delta, 3b_2^2, -b_2^3\right), \tag{34}$$

where $\delta$ is an integration constant. Thus, the exact solution for the radial geodesics of the LDBH results in

$$r = \frac{s}{x} + z = -z\sinh^2\left[\frac{\sqrt{b_2}}{2}(\gamma - c)\right]; \quad (z \leq 0), \tag{35}$$

or

$$r = \frac{s}{x} + z = z\left(\frac{3b_2}{12U_2 - b_2} + 1\right). \tag{36}$$

The significant cases about this solution are summarized below.

(*i*) If $\Xi = 0$, which means that $K = \varepsilon$ or $z = 0 \implies r(\gamma) = 0$.

(*ii*) If $\Upsilon = 0$, which means that $K = \alpha^2 r_o^2 + \varepsilon$ or $z \to -\infty \implies r(\gamma) \to \infty$.

On the other hand, one can easily integrate the $\theta$-equation (19) to obtain the following analytical solution:

$$\theta = \theta(\gamma) = \pi \pm \cos^{-1} \zeta, \tag{37}$$

by which

$$\zeta = \sqrt{\frac{K - \beta^2}{K}} \cos\left[\sqrt{K}(\gamma - \gamma_0)\right], \tag{38}$$

where $\gamma_0$ is yet another integration constant, and $K \geq \beta^2$ condition is imposed in order to have non-imaginary solution for $\theta(\gamma)$. After substituting the above solution into the $\phi$-equation (17), we find out

$$\phi = \phi(\gamma) = \phi_0 + \tan^{-1}\left\{\frac{\sqrt{K} \tan\left[\sqrt{K}(\gamma - \gamma_0)\right]}{\beta}\right\}, \tag{39}$$

where $\phi_0$ is also an integration constant. For the completeness of the geodesic equations, one should also investigate the solution for $t$-equation (16). If we substitute Eq. (35) into Eq. (16), and apply the transformation given below

$$v = \frac{\sqrt{b_2}}{4}(\gamma - c), \tag{40}$$

we obtain the following differential equation

$$\frac{dt}{dv} = \frac{4r_0^2 \alpha z \sinh^2(2v)}{\sqrt{b_2}\left[b + z \sinh^2(2v)\right]}. \tag{41}$$

With the aid of [28], the $t$-solution of Eq. (41) is obtained as

$$t = t(v) = \frac{2r_0^2 \alpha}{\sqrt{b_2}} \left[ ln\left(\frac{tanh\,v+1}{tanh\,v-1}\right) + b\frac{\Theta-z}{H\Theta}\left(tan^{-1}G - tanh^{-1}G\right) \right], \qquad (42)$$

where

$$\begin{aligned} \Theta &= \sqrt{z(z-b)}, \\ H &= \sqrt{b(\Theta+2z-b)}, \\ G &= \frac{b\,tanh\,v}{H}. \end{aligned} \qquad (43)$$

## 4. Conclusion

In this paper, we have considered geodesic structure of the LDBH, which is a solution to the EMD theory. Using the conventional Lagrangian procedure, the radial and angular EL equations for generic test particles have been obtained. The exact analytical solutions of the general radial geodesics with the Mino proper time have been given in terms of hyperbolic function (35) and the $\wp$-function (36). We have also represented the angular and time solutions as a function of the Mino proper time.

As a final remark, it would be interesting to extend our work to the geodesics of the rotating LDBHs [7,17,18]. By this way, we plan to investigate the effect of the rotation parameter on the geodesics of the LDBH. This is going to be our next study in the near future.

**Acknowledgement** We thank the anonymous referee and editor for their valuable and constructive suggestions**.**